\documentclass{emulateapj}

\usepackage{natbib}
\usepackage{multirow}
\begin{document}
\title{IRAS 20050+2720: Anatomy of a young stellar cluster}

\author{H.~M.~G\"unther, S.~J.~Wolk, B.~Spitzbart}
\affil{Harvard-Smithsonian Center for Astrophysics, 60 Garden Street, Cambridge, MA 02138, USA}
\email{hguenther@cfa.harvard.edu}
\and

\author{R.~A.~Gutermuth}
\affil{Department of Astronomy, University of Massachusetts, Amherst, MA 01003, USA}
\and

\author{J.~Forbrich, N.~J.~Wright}
\affil{Harvard-Smithsonian Center for Astrophysics, 60 Garden Street, Cambridge, MA 02138, USA}
\and

\author{L.~Allen}
%Lori Allen, NOAO <lallen@noao.edu>
\affil{National Optical Astronomy Observatory, 950 North Cherry Avenue, Tucson, AZ 85719, USA}
\and

\author{T.~L.~Bourke}
\affil{Harvard-Smithsonian Center for Astrophysics, 60 Garden Street, Cambridge, MA 02138, USA}
\and
 
%megeath@physics.utoledo.edu
\author{S.~T.~Megeath}
\affil{Department of Physics and Astronomy, MS-113, University of Toledo, 2801 W. Bancroft St., Toledo, OH 43606, USA}

\and
%jlpipher@pas.rochester.edu
\author{J.~L.~Pipher}
\affil{University of Rochester, 500 Wilson Boulevard, Rochester, NY 14627, USA}

\begin{abstract}
IRAS 20050+2720 is young star forming region at a distance of 700~pc without apparent high mass stars.
We present results of our multiwavelength study of IRAS 20050+2720 which includes observations by \emph{Chandra} and \emph{Spitzer}, and 2MASS and UBVRI photometry. In total, about 300 YSOs in different evolutionary stages are found. We characterize the distribution of young stellar objects (YSOs) in this region using a minimum spanning tree (MST) analysis. We newly identify a second cluster core, which consists mostly of class~II objects, about 10\arcmin{} from the center of the cloud. YSOs of earlier evolutionary stages are more clustered than more evolved objects.  The X-ray luminosity function (XLF) of IRAS 20050+2720 is roughly lognormal, but steeper than the XLF of the more massive Orion nebula complex. IRAS 20050+2720 shows a lower $N_H/A_K$ ratio compared with the diffuse ISM.
\end{abstract}


\section{Introduction}
Star formation occurs in dense and cool molecular clouds, which collapse under their own gravity. We observe those clouds over a range of masses, but the majority of stars forms in clusters with more than a hundred members \citep{2000AJ....120.3139C,2003ARA&A..41...57L,2003AJ....126.1916P}. Thus, studying star formation in those clusters is an important step to understand the history of the stars we currently observe, and the stellar population of the outer galaxy as a whole \citep[for general reviews of star formation see][]{2007prpl.conf.....R}.

In the early stages the proto-star is hidden by its parent circumstellar material and thus can only be observed as far-IR emission from the warm, dusty envelope (class 0). This envelope becomes less dense in class I objects. They often drive powerful outflows and carve holes in their envelopes so radiation can escape. Eventually, the envelope disperses and the star becomes visible as a classical T~Tauri star (CTTS) or ``class II'' source in IR classification \citep{1987IAUS..115....1L}. The disk still causes an IR excess over photospheric stellar emission. On the star itself accretion shocks and coronal activity lead to X-ray emission \citep{lamzin,2002ApJ...567..434K,acc_model,2011AN....332..448G}. Later, the IR excess vanishes as the gas disk is dissipated \citep{2012arXiv1205.2049I}. In this stage (weak-line T~Tauri stars = WTTS) cluster members cannot be distinguished from main-sequence (MS) stars by IR observations. One method to identify them is through their X-ray luminosity, which is far higher than for most older stars \citep[for a review of X-ray properties of young stellar objects (YSOs) before high-resolution X-ray spectroscopy with \emph{Chandra} or \emph{XMM-Newton} see][]{1999ARA&A..37..363F}.

%YSOs in dense clusters stars may influence each other during their evolution. Most notably strong winds or strong radiation fields of early-type stars can evaporate the disks of their late-type neighbors. In the Orion Nebula Cluster (ONC), this process can be directly observed in tadpole-shaped photo-evaporating proplyds \citep[e.g.][]{1994ApJ...436..194O,2012ApJ...746L..21W}.

Many details of the star formation process in clusters are still under debate despite significant observational progress. For example, it is unclear if in an undisturbed cluster high-mass or low-mass stars form first. However, the theoretical assumption of a cluster evolving with minimal influence from the rest of the galaxy may be far from realistic. For example, in $\rho$~Oph \citet{1992A&A...262..258D} suggested that the expanding shell of the upper Sco star forming region travelled through the cloud and triggered the star formation. 
Events like that would cause a non-uniform age distribution within a cluster. 
%If triggered star formation is the rule rather than the exception, it would be even more difficult to disentangle the competing processes that influence star formation.

In this paper we study the young cluster IRAS~20050+2720, which is located at a distance of 700~pc based on CO velocities \citep{1989ApJ...345..257W}.
It is found at a galactic longitude where the Cygnus Rift and the Cygnus X region are distinguishable, therefore, there is no distance ambiguity.  We use this distance throughout the article. There appear to be no massive stars in IRAS 20050+2720, thus the intensity of the ambient radiation should be small and we can study the evolution of late-type stars in the absence of external irradiation on their disks.

IRAS~20050+2720 is associated with a water maser \citep{1991A&A...246..249P} and a methanol maser \citep{2010A&A...517A..56F}. Molecular line emission indicates mass infall \citep{1997ApJ...484..256G,1999ApJS..122..519C}.
Radio observations reveal a complex structure of several lobes close to the nominal source position of IRAS~20050+2720, which has been identified as two or three jets from different proto-stellar sources \citep{1995ApJ...445L..51B,1999A&A...343..585C,2008A&A...481...93B}. %The fastest lobes contain knots that reach velocities up to 70~km~s$^{-1}$. 
The luminosity of this region is estimated as 388~$L_{\sun}$ from the \emph{IRAS} fluxes \citep{1996A&A...308..573M}. \citet{2001A&A...369..155C} obtained mm-maps of the regions, estimatinge a total gas mass of 200~$M_{\sun}$ within 65\arcsec{} (0.2~pc) radius.

% Diesen Satz habe ich irgendwo geklaut. Er enthalt weitere Referenzen, die ich z.T noch nicht verwendet habe: IRAS 20050+2720 is deeply embedded in a core that has been observed in the continuum at millimeter (Wilking et al. 1989; Chen et al. 1997; Choi et al. 1999; Chini et al. 2001; Furuya et al. 2005; Beltrán et al. 2006a) and centimeter (Wilking et al. 1989; Anglada et al. 1998) wavelengths and also in different high-density tracers, such as HCO+ and HCN (Choi et al. 1999), NH3 (Molinari et al. 1996), CS and N2H+ (Bachiller et al. 1995; Williams & Myers 1999), and 13CO and C18O (Ridge et al. 2003). 
% 1998ApJ...509..324D has 50 and 100 mu images and finds that IRAS XXX is an extended source - not very surprising. I think there is no need to cite that paper here.

The first attempt to establish an IR classification of individual stars in IRAS~20050+2720 identified about 100 cluster members, about half of which show an IR excess \citep{1997ApJ...475..163C}. While they estimate an average cluster age of 1~Myr, they interpret the radio lobes and the strong reddening as signatures for a more recent star formation event in the last 0.1~Myr. 
\citet{2005ApJ...632..397G} confirmed these ideas. %, noting one subcluster, that \citet{1997ApJ...475..163C} defined, lacks 850~$\mu$m emission which is present in the other two.

With \emph{Spitzer} it is possible to get very accurate photometry of star forming regions and classify individual objects according to their IR spectral energy distribution (SED). \citet{2009ApJS..184...18G} identified 177 YSOs in IRAS~20050+2720, which they grouped in two distinct cores. Compared with \citet{1997ApJ...475..163C} they identify more absorbed cluster members and show that the earlier definition of subclusters is partially due to an observational bias. In this paper, we extend the spatial coverage with an additional \emph{Spitzer} observation and add information from a \emph{Chandra} observation and, surprisingly, the first dedicated optical photometry.

Our goal is to compare the clustering and the X-ray properties of IRAS 20050+2720 with other star forming regions and to search for internal gradients of density and age that could indicate triggered star formation.

In Sect.~\ref{sect:obs} we describe the observations of IRAS~20050+2720. In Sect.~\ref{sect:analysis} we identify disk-bearing YSOs by the IR colors and disk-less YSOs based on their X-ray emission. Our results in Sect.~\ref{sect:results} describe the spatial distribution of YSOs and the optical photometry allows us to place individual members on evolutionary tracks. We show the X-ray properties and discuss the $A_K/N_H$ ratio. We end with a short summary in Sect.~\ref{sect:summary}.

\section{Observations}
\label{sect:obs}
In this section we describe the datasets we use for our analysis. They are summarized in table~\ref{tab:obslog}.

\begin{deluxetable}{lllrl}
\tablecaption{Log of observations\label{tab:obslog}}
\tablewidth{0pt}
\tablehead{
\colhead{Facility} & \colhead{Id} & \colhead{instrument} & \colhead{exposure} & \colhead{date}}

\startdata
Chandra & \dataset[ADS/Sa.CXO#obs/06438]{ObsId 6438} & ACIS-I & 23 ks & 2006-12-10\\
Chandra & \dataset[ADS/Sa.CXO#obs/07254]{ObsId 7254} & ACIS-I & 21 ks & 2006-01-07\\
Chandra & \dataset[ADS/Sa.CXO#obs/08492]{ObsId 8492} & ACIS-I & 51 ks & 2007-01-29\\
Spitzer & AOR 3656448  & IRAC & 41.6 s & 2004-06-09\\
Spitzer & AOR 3665152  & MIPS & 40.4 s & 2004-09-25\\
Spitzer & AOR 21893120 & IRAC & 41.6 s & 2007-06-29\\
Spitzer & AOR 21893376 & IRAC & 41.6 s & 2007-06-29\\
Spitzer & AOR 21891840 & MIPS & 40.4 s & 2007-10-27\\
FLWO    & filter: $B$  & Keplercam & 6-300 s &  2009-11-18\\
FLWO    & filter: $V$  & Keplercam & 6-300 s & 2009-11-18\\
FLWO    & filter: $r'$ & Keplercam & 6-300 s & 2009-11-18\\
FLWO    & filter: $i'$ & Keplercam & 6-300 s & 2009-11-18\\
FLWO    & filter: $u'$ & Keplercam & 6-300 s & 2009-11-21\\
FLWO    & filter: $B$  & Keplercam & 6-300 s & 2009-11-21\\
FLWO    & filter: $V$  & Keplercam & 6-300 s & 2009-11-21\\
FLWO    & filter: $r'$ & Keplercam & 6-300 s & 2009-11-21\\
FLWO    & filter: $i'$ & Keplercam & 6-300 s & 2009-11-21\\
\enddata

\end{deluxetable}

\subsection{Spitzer}
\label{sect:spitzer}
%Observationally, IRAS~20050+2720 is especially suited for our study, because the diffuse background at 24~$\mu$m is weak.
\emph{Spitzer} has observed this region several times. We use data from two initial observations performed in 2004, which are described in detail in \citet{2009ApJS..184...18G}. This dataset misses an extended population beyond the cluster core, so follow-up observations were performed in 2007 as part of the \emph{Spitzer} ``Clusters Near Clusters'' GTO program (Program ID 40147) \citep{2012arXiv1204.3110M}. In 2004 the cluster center was imaged with IRAC \citep[$3.6-8.0\;\mu$m,][]{ 2004ApJS..154...10F} with a $3\times4$ raster map of four dithered images per position in the high dynamic range mode. In 2007 two additional $3\times4$ raster maps were obtained to the north-east and south-west of the original pointing, which overlap partially with the initial observation. Table~\ref{tab:obslog} gives the exposure time per pixel for each observation. MIPS mosaics \citep[$24-160\;\mu$m,][]{2004ApJS..154...25R} were taken at medium scan rate with full width scan stepping. The observation in 2004 contains three scanning stripes, one stripe towards the south-east was added in 2007. Because of the low angular resolution the 70 and 160~\micron{} data cannot be used for individual sources in a dense region like IRAS~20050+2720. 
All datasets are reduced and combined following  \citet{2009ApJS..184...18G} using the \texttt{ClusterGrinder} IDL package, which is described in detail by \citet{2008ApJ...674..336G}. In short, the standard basic calibrated data products (BCD) from the \emph{Spitzer} Science center pipeline are treated for bright source artifacts and combined into a mosaic image, eliminating outliers to remove cosmic rays. The mosaic of individual AORs are shifted to match the 2MASS astrometry. Automated source detection and aperture photometry are carried out on the mosaics.

Compared with the reduction described by \citet{2009ApJS..184...18G} our new data reduction is improved in several ways to increase the photometric accuracy \citep{2012arXiv1204.3110M}. Our data are now processed with BCD S18.5 which contains a muxbleed correction, resulting in a cleaner mosaic close to the bright sources in the center of the cluster. Also, we now use a pixel scale of 0.86\arcsec{}, which is  $1/\sqrt{2}$ of the native pixel size. This finer grid gives better centroid positions; it also allows to better separate close sources. Sect.~\ref{sect:improvedspitzer} describes how the improved reductions effect the detection efficiency.

\subsection{Chandra}
\label{sect:chandra}
IRAS~20050+2720 was observed with \emph{Chandra} using the VFAINT mode of the ACIS-I detector. As indicated in table~\ref{tab:obslog} the cluster was observed in three different epochs. The position angle varies between observations, thus the coverage of the field is non-uniform on the edges.
The individual exposures are merged and processed with the pipeline of the ANCHORS project \citep{2005prpl.conf.8518S}.
% Background is suppressed using the VFAINT mode, if reprocessed in VFAINT mode ...
Source detection is performed with the \texttt{wavdetect} algorithm in CIAO \citep{2006SPIE.6270E..60F}. We keep sources with a significance above $2\sigma$.
For each source with more than 20 counts, event lists and response files are generated. 
We fit a single Raymond-Smith thermal emission model \citep{2010AAS...21547607F}, where the absorbing column density $N_H$, the temperature $T$ and the volume emission meassure are free parameters. The metalicity is fixed at 0.3 compared with solar \citep{1989GeCoA..53..197A}. Also, for each source a lightcurve is generated and its Gregory-Loredo \citep{1992ApJ...398..146G} statistic is calculated to characterize the source variability but this will be discussed in a later paper. The spectral properties presented here are effectively time-averaged over all three observations.

In total 368 sources are detected with \emph{Chandra}. Table~\ref{tab:xray} lists the position of the X-ray sources and their properties ($f_X$, $N_H$ and k$T$ and observed flux  in the energy range 0.3-8.0~keV) with statistical errors. Also, we calculated the unabsorbed luminosities assuming a distance of 700~pc. In addition, this table contains the ID number of the source in the merged source table (table~\ref{sourcelist}) and the classification (see Sect.~\ref{sect:classification}-\ref{sect:xraysources}). Table~\ref{tab:xray} is ordered by the ID number from the merged source table, which in turn is ordered by the classification.

\begin{table*}
\caption{X-ray sources and spectral fits \label{tab:xray}}
\begin{tabular}{rrrrrrrrrrrrrrr}
\hline \hline
ID & $\alpha$ & $\delta$ & class & net counts & $f_X$ & $\sigma_{f_X}$ & $N_H$ & $\sigma_{N_H}$ & k$T$ & $\sigma_{\textnormal{k}T}$ & obs. flux & $\sigma_{obs. flux}$ & $\log L_X$ \\
 & [\degr] & [\degr] & & & \multicolumn{2}{c}{[$10^{-7}$~s$^{-1}$~cm$^{-2}$]} & \multicolumn{2}{c}{[$10^{22}$~cm$^{-2}$]} & \multicolumn{2}{c}{[keV]} & \multicolumn{2}{c}{[$10^{-15}$ erg s$^{-1}$ cm$^{-2}$]} & [log(erg s$^{-1}$)]\\  \hline
1 & 301.7633 & 27.4821 & XYSO &    61.5 &  31 &   8 & 0.6 & 0.2 & 4.2 & 0.0 & 7.8 & 1.4 & 29.9 \\
3 & 301.7693 & 27.4823 & XYSO &    23.6 &  10 &   4 & 10.5 & 1.8 & 0.7 & 0.1 & 4.5 & 1.9 & 31.7 \\
11 & 301.7786 & 27.4701 & XYSO &    40.9 &  18 &   5 & 2.4 & 0.6 & 6.0 & 1.9 & 8.9 & 2.0 & 30.0 \\
12 & 301.7796 & 27.4808 & XYSO &    47.2 &  21 &   6 & 18.6 & 5.8 & 3.7 & 1.9 & 17.2 & 14.0 & 30.8 \\
13 & 301.7803 & 27.4834 & XYSO &    32.1 &  18 &   6 & 2.0 & 0.4 & 2.7 & 0.2 & 5.5 & 1.2 & 30.0 \\
69 & 301.7722 & 27.5010 & I &    26.6 &  12 &   5 & 4.2 & 1.2 & 3.0 & 1.5 & 6.2 & 3.5 & 30.1 \\
73 & 301.7772 & 27.4814 & I &    32.8 &  17 &   6 & 2.4 & 0.7 & 4.2 & 1.4 & 5.6 & 1.5 & 29.9 \\
131 & 301.7531 & 27.4996 & II &    64.9 &  36 &   8 & 1.2 & 0.4 & 1.3 & 0.3 & 5.6 & 2.5 & 30.1 \\
132 & 301.7535 & 27.5658 & II &    56.2 &  31 &   9 & 0.4 & 0.2 & 3.6 & 0.0 & 7.5 & 1.7 & 29.8 \\
156 & 301.7638 & 27.4985 & II &   328.4 & 250 &  18 & 0.7 & 0.1 & 5.5 & 0.8 & 56.6 & 4.8 & 30.7 \\
160 & 301.7658 & 27.5122 & II &    21.9 &  14 &   5 & 0.1 & 0.2 & 12.0 & 0.0 & 2.9 & 0.8 & 29.3 \\
170 & 301.7706 & 27.4822 & II &    47.7 &  21 &   6 & 2.9 & 0.7 & 1.7 & 0.5 & 7.3 & 3.6 & 30.3 \\
174 & 301.7723 & 27.4846 & II &    61.8 &  41 &   9 & 1.2 & 0.3 & 5.9 & 0.0 & 9.1 & 1.7 & 30.0 \\
177 & 301.7735 & 27.4925 & II &    14.9 &  10 &   6 & 1.4 & 0.6 & 3.1 & 0.1 & 2.5 & 1.0 & 29.5 \\
181 & 301.7753 & 27.4791 & II &    43.4 &  27 &   7 & 2.8 & 0.8 & 1.8 & 0.6 & 7.0 & 4.0 & 30.3 \\
183 & 301.7774 & 27.4936 & II &    99.2 &  54 &  10 & 1.4 & 0.3 & 3.0 & 0.9 & 13.4 & 3.8 & 30.3 \\
187 & 301.7784 & 27.4542 & II &    41.7 &  23 &   7 & 3.2 & 1.1 & 1.8 & 0.7 & 6.2 & 4.7 & 30.3 \\
214 & 301.8466 & 27.5287 & II &    17.7 &  10 &   5 & 0.9 & 0.4 & 4.0 & 0.1 & 2.3 & 0.8 & 29.4 \\
275 & 301.7790 & 27.4958 & II* &   492.9 & 274 &  22 & 1.1 & 0.1 & 3.2 & 0.5 & 60.7 & 7.4 & 30.9 \\
286 & 301.7363 & 27.5443 & III &    67.1 &  40 &   9 & 0.7 & 0.2 & 4.0 & 1.1 & 8.3 & 1.7 & 29.9 \\
287 & 301.7386 & 27.3951 & III &    56.8 &  33 &   9 & 0.0 & 0.0 & 0.8 & 0.1 & 5.2 & 0.9 & 29.5 \\
289 & 301.7556 & 27.4678 & III &    15.2 &   8 &   3 & 2.6 & 0.8 & 0.7 & 0.2 & 1.8 & 1.8 & 30.5 \\
292 & 301.7678 & 27.4655 & III &    39.3 &  21 &   6 & 4.6 & 1.2 & 1.6 & 0.5 & 6.2 & 4.1 & 30.4 \\
293 & 301.7685 & 27.4901 & III &    55.5 &  36 &   7 & 0.7 & 0.2 & 2.5 & 0.9 & 5.7 & 1.6 & 29.8 \\
295 & 301.7730 & 27.4579 & III &   164.9 & 104 &  14 & 2.7 & 0.3 & 3.2 & 0.5 & 35.6 & 5.7 & 30.8 \\
310 & 301.8392 & 27.5042 & III &   561.9 & 332 &  25 & 0.7 & 0.1 & 2.9 & 0.3 & 62.9 & 6.0 & 30.9 \\
312 & 301.8469 & 27.4258 & III &   108.4 &  46 &  11 & 0.9 & 0.2 & 3.6 & 0.0 & 17.7 & 2.3 & 30.3 \\
313 & 301.8526 & 27.5512 & III &    53.9 &  35 &   9 & 0.2 & 0.2 & 0.3 & 0.1 & 3.8 & 4.1 & 29.8 \\
316 & 301.8716 & 27.4768 & III &    30.6 &  22 &   7 & 0.8 & 0.3 & 4.6 & 0.0 & 5.9 & 1.6 & 29.8 \\
319 & 301.9011 & 27.5514 & III &    62.5 &  41 &  12 & 1.8 & 0.4 & 1.0 & 0.2 & 5.6 & 2.9 & 30.4 \\
325 & 301.9235 & 27.5355 & III &   133.3 &  64 &  12 & 0.6 & 0.1 & 2.0 & 0.2 & 15.3 & 2.2 & 30.3 \\
327 & 301.9519 & 27.4290 & III &   322.6 & 203 &  24 & 0.0 & 0.0 & 1.0 & 0.0 & 31.1 & 2.0 & 30.3 \\
\hline
\end{tabular}
\tablecomments{This table shows only X-ray sources, that are identified as YSOs and have valid spectral fits. The full table including all detected X-ray sources can be found in the electronic version.}
\end{table*}

\begin{deluxetable}{rlrrc}
\tablecaption{Format of electronic table\label{sourcelist}}
\tablehead{\colhead{ID} & \colhead{name} & \colhead{unit} & \colhead{symbol} & \colhead{comments}}
\startdata
1 & ID &  & ID &  \\
2 & RA & deg & $\alpha$ & J2000 \\
3 & DEC & deg & $\delta$ & J2000 \\
4 & U & mag & $U$ & FLWO \\
5 & U\_ERR & mag & $\sigma_{U}$ & FLWO \\
6 & B & mag & $B$ & FLWO \\
7 & B\_ERR & mag & $\sigma_{B}$ & FLWO \\
8 & V & mag & $V$ & FLWO \\
9 & V\_ERR & mag & $\sigma_{V}$ & FLWO \\
10 & R & mag & $R$ & FLWO \\
11 & R\_ERR & mag & $\sigma_{R}$ & FLWO \\
12 & I & mag & $I$ & FLWO \\
13 & I\_ERR & mag & $\sigma_{I}$ & FLWO \\
14 & J & mag & $J$ & 2MASS \\
15 & J\_ERR & mag & $\sigma_{J}$ & 2MASS \\
16 & H & mag & $H$ & 2MASS \\
17 & H\_ERR & mag & $\sigma_{H}$ & 2MASS \\
18 & K & mag & $K$ & 2MASS \\
19 & K\_ERR & mag & $\sigma_{K}$ & 2MASS \\
20 & IRAC\_1 & mag & $[3.6\mu]$ & \emph{Spitzer} \\
21 & IRAC\_1\_ERR & mag & $\sigma_{[3.6\mu]}$ & \emph{Spitzer} \\
22 & IRAC\_2 & mag & $[4.5\mu]$ & \emph{Spitzer} \\
23 & IRAC\_2\_ERR & mag & $\sigma_{[4.5\mu]}$ & \emph{Spitzer} \\
24 & IRAC\_3 & mag & $[5.8\mu]$ & \emph{Spitzer} \\
25 & IRAC\_3\_ERR & mag & $\sigma_{[5.8\mu]}$ & \emph{Spitzer} \\
26 & IRAC\_4 & mag & $[8.0\mu]$ & \emph{Spitzer} \\
27 & IRAC\_4\_ERR & mag & $\sigma_{[8.0\mu]}$ & \emph{Spitzer} \\
28 & MIPS & mag & $[24\mu]$ & \emph{Spitzer} \\
29 & MIPS\_ERR & mag & $\sigma_{[24\mu]}$ & \emph{Spitzer} \\
30 & Ha & mag & $H\alpha$ & IPHAS \\
31 & Ha\_ERR & mag & $\sigma_{H\alpha}$ & IPHAS \\
32 & r & mag & $r$ & IPHAS \\
33 & r\_ERR & mag & $\sigma_{r}$ & IPHAS \\
34 & i & mag & $i$ & IPHAS \\
35 & i\_ERR & mag & $\sigma_{i}$ & IPHAS \\
36 & class &  & class &  \\
37 & AK & mag & $A_K$ & see Sect~\ref{sect:classification} \\
38 & AV & mag & $A_V$ & from extinction map \\
\enddata
\tablecomments{The data for this table is published in the online version of the journal only.}
\end{deluxetable}

\subsection{Optical photometry}
Optical photometry of IRAS~20050+2720 was obtained at the Fred Lawrence Whipple Observatory (FLWO) with the 1.2~m telescope and the Keplercam instrument. Keplercam is a camera with a Fairchild ``CCD~486'' detector. 
%Each pixel corresponds to 0.336\arcsec on the sky and saturates at about 65,00 counts per pixel with a total field-of-view (FOV) $23.1\times23.1$\arcmin. We operated the chip in the default $2\times2$ binning. It is read-out in four amplifiers, each of them has a different gain and read-out noise. 
Observations were taken with staggered exposure times of 6, 60 and 300~s, so that photometry for brighter stars can be performed on frames with shorter exposure times. Keplercam is equipped with filters for different photometric systems, here we use observations taken in $u'B_H V_H r'i'$. The $B_H$ and $V_H$ filters belong to the Harris set, the $u'$, $r'$ and $i'$ filters are part of a Sloan Digital Sky Survey (SDSS) set \citep{1996AJ....111.1748F}.

We performed the data reduction in Pyraf, a Python interface to IRAF \citep{1993ASPC...52..173T}. Photometry is done with PSF-fitting in DAOPHOT \citep{1987PASP...99..191S} as implemented in IRAF.
Data and flatfield are bias-corrected, then the data images are divided by the flatfield. %On those images we performed a source detection and an initial aperture photometry as starting point for the PSF-fitting. The PSF is fitted independently for each exposure and each amplifier, but is assumed to not to vary spatially. We approximate the PSF with a \texttt{moffat25} function and calculate an look-up table of deviation from this function from the brightest, non-saturated sources in the field. We place a very conservative limit of 40,000~counts per pixel to ensure linearity of the detector. The full width at half maximum (FWHM) of the PSF is about 4.5 binned pixel, i.e. about 3\arcsec, but the PSF changes from exposrue to exposure and in some cases it is significantly non-symmetric. This leads to notable artifacts in PSF-subtracted images and limits or ability perform a second source detection. Also, the tail of the PSF is very wide an merges into the background after 90~binned pixels or more for bright sources. Thus, our source lists are incomplete close to bright sources and therefore the completeness limit is non-uniform over the FOV.

The absolute pointing accuracy of the FLWO telescope is insufficient to match source positions with the other datasets directly. Therefore, we first correct the WCS of each individual exposure with the IRAF task \texttt{msccmatch}, which minimizes the residuals between sources in our images and the 2MASS catalog \citep{2006AJ....131.1163S}. Sources are extracted independently for different exposures and source lists are merged with a matching tolerance of 3\arcsec, which is of the order of the remaining astrometric errors. Properties of bright sources are taken from shorter exposures to avoid the non-linear regime of the detector. %, which are identified in exposures with different exposure times, are taken from the shorter exposure. %, if their instrumental magnitude is above a threshold value.

To perform absolute photometry extinction and zero point of the instrumental magnitude are calibrated with observations of Landolt field~01 \citep{2000PASP..112..925S}, which was reduced analogues to the data of IRAS~20050+2720 itself. We fit the instrumental magnitudes in $u'B_H V_H r'i'$ to the catalog values for $UBVRI$ assuming an offset, a linear color term and a linear extinction term. Ten sources in the standard field are extremely red ($1.15 > R-I > 2.0$), but we do not find the break in the transformation law indicated in \citet{2002AJ....123.2121S} for very red sources. Instead the full sample fits the relation $R-I \propto r'-i'$ within the observational uncertainties. We remove the faintest stars from our final sourcelist, so that the typical error on the flux is about 10\%.

The optical photometry is supplemented by data from IPHAS, the Isaac Newton Telescope (INT) Photometric H$\alpha$ Survey \citep{2005MNRAS.362..753D}. IPHAS is a survey of the northern Galactic Plane in broad-band $r'$, $i'$, and narrow-band H$\alpha$ filters to a depth of $r' \simeq 20$ (10$\sigma$). Using the IPHAS Initial Data Release \citep{2008MNRAS.388...89G} we extracted all stellar sources (morphological classification = -1) within 15$^\prime$ of the center of IRAS 20050+2720. Because IPHAS images the sky twice (with pointings offset to cover the Wide Field Camera chip gaps) most sources are observed twice and so these were cross-matched and the mean magnitudes calculated. We remove all sources with photometric uncertainties in any band above 0.1~mag. This results in a total of $\sim$1200 sources.

% \subsection{VLA}
% IRAS~20050+2720 was observed by the VLA several times. Here, we analyze program ID~AA0188, which has the longest continuous on-source-time ($3\times1200$~s). IRAS~20050+2720 was observed in A configuration at 8.4~GHz.
% The phase center was $(\alpha, \delta)_{ {\rm J}2000}= 20^{\rm h}05^{\rm m}03^{\rm s},+27^\circ20'09''$. The field of view is much smaller than for all other bands. We reconstruct an image with a size of $\approx1$\arcsec{}. Data are analyzed using the NRAO Astronomical Image Processing System (AIPS).
% \object{4C 31.56} was observed as a calibrator bracketing the observations of IRAS~20050+2720.

\section{Data analysis}
\label{sect:analysis}
\subsection{Source matching}
The different infrared observations are matched as in \citet{2008ApJ...674..336G}. Uncertainties on the coordinates are 0\farcs25 ($1\sigma$). We then add the X-ray data, matching X-ray sources and IR sources, if they are closer than the large axis of the $3\sigma$ error ellipse from \texttt{wavdetect}. If more then one IR source fullfills this criterion, we match the closest X-ray and IR-source (less than 5 cases). If two X-ray sources are matched to the same IR source we keep the match, which is closer to the IR centroid and list the remaining X-ray source as unmatched (1 case).
Optical sources are identified with IR or X-ray sources using a constant matching radius of 3\arcsec.

The final list of cross-matched sources is published online as an electronic table, table~\ref{sourcelist} describes its content. This table also contains a classification (see Sect~\ref{sect:classification}-\ref{sect:xraysources}) and an estimate of the total optical extinction $A_V$ of the cloud at the source position from the extinction map of \citet{2009ApJS..184...18G}. This map is based on $H-K$ colors of 2MASS sources, assuming an average intrinsic color of $(H-K)=0.2$ \citep{2005ApJ...632..397G}.

%The radio source positions are accurate to 0\farcs3, thus we match them to the IR sources with a matching radius of 0\farcs4. Only three IR sources are detected in the radio band, these have the ID numbers 4, 274 and 20516 in table~\ref{sourcelist}.

\subsection{Class I and II sources}
\label{sect:classification}
Class I and II sources are surrounded by circumstellar material, thus they have a distinct IR signature. We use the same classification method as in \citet{2009ApJS..184...18G}, which is based on a series of cuts in color-color diagrams (CCD) and color-magnitude diagrams (CMD) to remove various contaminants (galaxies with PAH emission, AGN, Herbig-Haro objects). The classification is primarily based on IRAC fluxes; however, if either the 5.8~\micron{} or the 8.0~\micron{} flux is not available, then the 2MASS bands can be used to estimate $A_K$ and deredden the remaining IRAC data to perform the classification. The dereddening is based on the extinction law from \citet{2007ApJ...663.1069F} and the reddening loci from \citet{1997AJ....114..288M}. In a final step, sources which are highly reddened and thus invisible at short wavelengths, but bright in the MIPS 24~\micron{} band are added. Likely, they are deeply embedded protostars, but their evolutionary status cannot be confirmed without submillimeter data. In the tables they are denoted as class~I* to distinguish them from less embedded class~I sources.
Sources with less surrounding material but an IR excess in all bands due to a disk are classified as class~II; those where the IRAC bands follow the energy distribution of a stellar photosphere, but an excess is found in MIPS 24~\micron{} are marked as class~II*. These sources are sometimes called transition objects because an excess is present only at longer wavelength and therefore larger disk radii. In an evolutionary sequence this can be interpreted as a transition phase between disk-bearing class~II objects and MS stars. Sources compatible with stellar photospheres are marked as class~III candidates. These could either be young pre-main sequence WTTS or main-sequence (MS) stars and are further classified based on optical and X-ray properties. Table~\ref{tab:sources} gives the total number of detected IR sources. The dataset contains 57 class~I sources, about 20\% of them (10) are detected in X-rays, and 183 class~II sources, 30\% (53) of them are detected in X-rays.

\begin{table}
\begin{center}
\caption{Number of IR sources\label{tab:sources}}
\begin{tabular}{lrr}
\tableline\tableline
Type & IR & X-ray\\
\tableline
XYSO (see text) & 0 & 18\\ 
Class I*  &    13 &     1\\ 
Class I   &    57 &    10\\ 
Class II  &   183 &    53\\ 
Class II* &     5 &     4\\ 
WTTS$^a$ & \multirow{2}{*}{2489${\bigl \lbrace}$} & 54\\ 
photosphere$^a$ &  & 9$^b$\\ 
\tableline
\end{tabular}
\end{center}
$^a$:~WTTS and MS stars both show stellar photospheric colors in the IR. $^b$:~Foreground stars.
\end{table}

\subsection{Improved Spitzer data reduction}
\label{sect:improvedspitzer}
In Sect.~\ref{sect:spitzer} \citep[see also][]{2012arXiv1204.3110M} we described several changes in our data reduction compared with \citet{2009ApJS..184...18G}. The improved photometry allows us to detect additional YSOs and in some cases objects are reclassified. The number of detected class~I sources rises from 41 to 51 in the central part of the 2004 observations. This field is not covered by the new observations, thus, the higher number is entirely due to an improved data treatment. The number of class~II sources rises from 95 to 105 and the number of deeply embedded sources doubles from 6 to 11. In both cases 2 sources are classified as transition disks (class II*).

% Class:   total    | orig area
%        2009  now  | 2009  now
% I    :   46   57 |    41   51
% II   :  118  183 |    95  105
% I*   :    7   13 |     6   11
% II*  :    6    5 |     2    2

\subsection{Class III sources}
\label{sect:xraysources}

The main goal of this study is to extend the treatment of YSOs to include young disk-less cluster members. While class~I and class~II objects can be identified by their IR excess, class~III sources have lost their envelope and their disk. In the IR they are indistinguishable from MS-stars. However, because of their low age they are still fast rotators and consequently luminous X-ray emitters. Our set of X-ray sources which do coincide with an IR source consistent with stellar photospheric IR colors, thus provides a list of strong candidates. All IR sources consistent with stellar photospheric IR colors but without an X-ray counterpart are reclassified as (main-sequence) ``star'' in table~\ref{sourcelist}.

 %The detection fraction of class~II sources is higer, because more evolved sources have in general less circumstellar absoprtion.

\subsubsection{Optical data}
% Most of the YSOs in IRAS~20050+2720 are not detected in our optical dataset. The limiting magnitude is about 19 in the $R$-bad, which is insufficient for late-type objects in 700~pc, if they are reddened by a few magnitudes of dust in the proto-stellar cloud. The detection efficiency is much better in the IR, where the dust opacity is lower. In Fig.~\ref{fig:CMD} (left panel) we show the $J/J-H$ color-magnitude diagram (CMD) of our dataset. We also an isochrone for 1~Myr \citep{2000A&A...358..593S} and a reddening vector for $A_V = 5$ \citep{1998ApJ...500..525S}. Both represent typical values for IRAS 20050+2720 and will be used throughout the paper. There is very little overlap between the optical sources and IR class~II sources or even the class~III candidates. We can make use of this fact to identify X-ray emitting foreground object to remove them from the sample of potential class~III sources.

\begin{figure}
\plotone{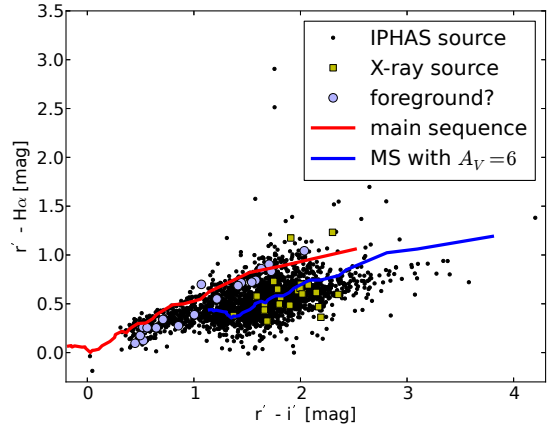}
\caption{Color-color diagram of all IPHAS sources within 15\arcmin{} of the center of the cluster. X-ray sources are marked with squares. Overplotted is an unreddened and a reddened MS from \citet{2005MNRAS.362..753D}.
See the electronic edition of the Journal for a color version 
of this figure.\label{fig:IPHAS}}
\end{figure}

Different methods can be applied to identify foreground sources in the sample. The three brightest stars are spectroscopically classified by \citet{1995A&AS..110..367N}, thus their distance is easy to determine. Beyond that we rely on the IPHAS catalog and our own photometry and use a procedure similar to \citet{2010ApJ...713..871W}. Fig.~\ref{fig:IPHAS} shows the IPHAS CCD for all sources within 15\arcmin{} of the cluster center. Overplotted are all X-ray detections. 

Class~II stars and young class~III objects are typically $H\alpha$ emitters, thus they are found above the MS. Giants and AGB stars are also found in this region, the latter extending towards much redder colors \citep{2008MNRAS.390..929W}. There are eleven sources with $r'-H\alpha$ significantly above the MS (more than three times the measurement uncertainty, but at least 0.2~mag) in figure~\ref{fig:IPHAS}. We exclude the object at $r'-H\alpha=1.5$ and $r'-i'=4$ which probably is an AGB star \citep{2008MNRAS.390..929W}. Four of the remaining ten sources are outside the FOV of IRAC bands 1 and 3 (ID 23216, 25132, 25423, 27352 in table~\ref{sourcelist}), one is undetected in IRAC band 3 and 4 (ID 12295) and thus those sources do not have an IR classification. Four sources coincide with class~II sources (ID 92, 100, 105, 146). The remaining source is IR-classified as a star (ID 1645).

\begin{figure}
\plotone{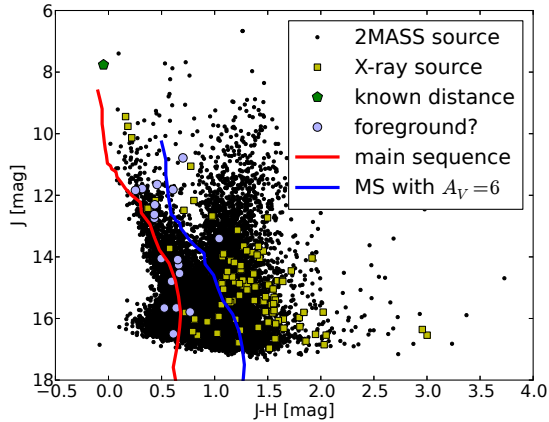}
\caption{J vs. J-H color-magnitude diagram. Filled colored symbols show X-ray sources with known parallax (green pentagon), those which are close to the MS in Fig.~\ref{fig:IPHAS} and thus are potential foreground sources (red circles) and other X-ray sources (yellow squares).
The early MS from \citet{2000A&A...358..593S} is overplotted for a distance of 700~pc. % The arrow indicates the reddening vector according to the compilation of \citet{1998ApJ...500..525S}.
See the electronic edition of the Journal for a color version 
of this figure.\label{fig:JJH}}
\end{figure}

Sources which are reddened by dust are located below and to the right of the MS, thus we expect cluster members and background sources in this region. Foreground objects should cluster around the MS; however, reddened young stars with H$\alpha$ in emission could also occupy this region, e.g. a young star with $A_V = 6$ and $H\alpha$ emission such that $r'-H\alpha$ is about 0.3~mag higher than for a MS star would be very close to the unreddened MS in figure~\ref{fig:IPHAS}. In this step we take all X-ray sources within 0.2~mag in either color of the MS as potential foreground sources. Next we plot those sources on a $J$ vs. $J-H$ color-magnitude-diagram in Fig.~\ref{fig:JJH} and compare their position with a MS at 700~pc. In this diagram reddening moves sources of intermediate mass roughly in parallel to the MS, thus the moderate reddening experienced by foreground stars does not shift them into the region where the YSOs of IRAS 20050+2720 are located. Given the spread of all IR sources in the IPHAS CCD, it seems very unlikely that a significant fraction of YSOs with H$\alpha$ emission coincides by chance with the unreddened MS in figure~\ref{fig:IPHAS} and also with the MS on the $J$ vs. $J-H$ diagram. Thus we remove nine sources which are compatible with unreddened MS stars at distances between 100 and 700~pc on both diagrams from our list of young class~III candidates.

% 
% Fig.~\ref{fig:CMD} (right panel) shows the CMD of all optically detected sources (black dots). Sources with X-ray emission are marked with a blue triangle, class~II sources with a green square. For illustrative purposes the pre-main sequence isochrone for a cluster of 1~Myr is also shown, reddened by $A_V=5$. This corresponds to the age of IRAS~20050+2720 \citep{1997ApJ...475..163C} and a typical value for the thickness of the cloud \citep[][Fig. 2]{2008ApJ...674..336G}.
% 
% The isochrone matches the class~II sources reasonably well, but there are about a dozen X-ray sources to the left of the isochrone in the diagram. In this $R/R-I$ diagram the extinction moves stars basically in parallel to the isochrone, thus the bluer colors cannot be explained with lower extinction values; sources in this this region must be foreground objects, which can be confirmed by overplotting the MS for distances of e.g. 300~pc.
% 
% In order to reduce the contamination with foreground objects we remove X-ray souces with optical colors $R-I < 0.95$, that are above the MS at 700~pc in the $R/R-I$ diagram. This excludes 11 X-ray sources from our list of class~III candidates. The color cut correspondes to unreddened stars of spectral types earlier than M.

\subsubsection{Other X-ray sources}

\begin{figure}
\plotone{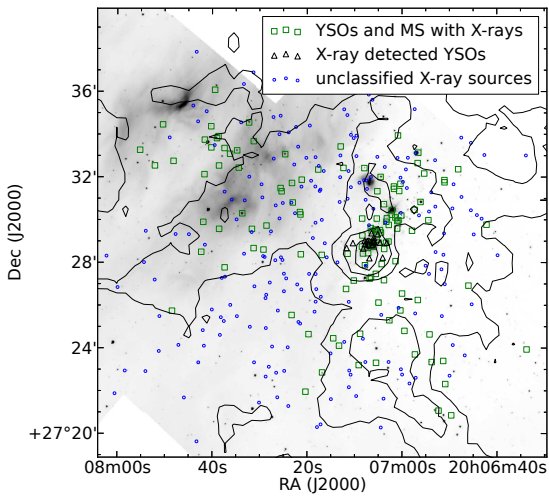}
\caption{Distribution of X-ray sources. The contours show the extinction from a background extinction map. The countour levels are $A_V=2,6,10,14$~mag after subtracting a background of $A_V=2$ \citep{2009ApJS..184...18G}. The gray scale image shows IRAC 8~\micron{} emission.
See the electronic edition of the Journal for a color version 
of this figure.\label{fig:X-raydist}}
\end{figure}

MS stars in front of the cloud are detected in the optical and IR down to spectral type M. Thus, the remaining unidentified sources are mostly background objects, which should be randomly distributed over the \emph{Chandra} field-of-view.  Fewer X-ray sources are expected towards the edges because the effective area of the ACIS detector is lower here than in the center. Additionally, we expect fewer background X-ray sources where the cluster is densest, because the extinction hides weak background sources. Figure~\ref{fig:X-raydist} shows the distribution of all X-ray sources.

Squares mark all sources which have been classified above, i.e.   they belong to one of the following groups: I*, I, II, II*, young class~III or MS. Small circles and triangles mark all as yet unidentified sources. They could be background sources like AGN, high-mass X-ray binaries etc., but this group can also include cluster members with insufficient IR data for an IR classification, either because they are highly obscured or due to source confusion in the densest part of the cluster. As expected these sources are  distributed across the field, but a close cluster is seen in the center of the cloud (triangles), where $A_V>10$~mag. Given the low density of background sources in the rest of the field, this is statistically a highly significant clustering. It is very unlikely to find a clustered group of unrelated background sources such as AGN exactly behind the densest part of IRAS 20050+2720. Thus, we assume that (most of) these sources must be young stars. Only one source in the group is detected in IPHAS, but without excess $H\alpha$ emission. 8 out of 18 sources have no IR counterpart at all, 10 are detected in one or more IR bands. Mostly the information is insufficient for the classification scheme used above. The remaining two sources are classified as stars with contamination by diffuse PAH emission in the aperture. They are otherwise compatible with a class~I or II source. All objects in this group are brighter for longer wavelengths for all bands where they are detected. This is compatible with (reddened) YSOs, but their exact status cannot be confirmed with the available data. In the following, X-ray sources in this region ($301.76<\alpha<301.81$ and $27.47<\delta<27.49$ - triangles in Fig.~\ref{fig:X-raydist}) are marked as \emph{X-ray (detected) YSOs = XYSO} and included in the YSO sample in table~\ref{sourcelist} and \ref{tab:sources}. The cutoff at $A_V>10$~mag is chosen to include the  group of X-ray sources in the center of the cluster. It is possible that individual YSOs at lower values of $A_V$ are missed, however the only way to identify them without sufficient optical or IR information is as a spatial overdensity compared with the background.

Hard X-rays, like radio radiation, penetrate much higher absorbing column densities than optical or IR photons and therefore it is not unusual to detect sources in the densest cores of young clusters only in these bands \citep[e.g.][for Cep~A, a massive star forming region at 730~pc, a distance very similar to IRAS 20050+2720]{2001ApJ...563..919H,2007A&A...469..207C,2009A&A...508..321S,2009ApJ...704.1495P}. 

\subsection{Sample completeness}
\label{sect:completness}
%We compare the X-ray flux distribution to a simulated distribution from the galactic model of \citet{2010arXiv1012.0314W}. The two distributions match closely for fluxes $>2\times 10^{-14}$~erg~s$^{-1}$~cm$^{-2}$, below this value we detect a higher number of sources than predicted due to the presence of IRAS 20050+2720. However, the cluster material absorbs background sources, which is not taken into account in the simulations. Thus, the difference between the observed and the simulated distribution only gives a lower limit to the number of cluster members of $\approx 20$, much less than the number of YSOs found in X-rays. Our detection efficiency drops for lower fluxes. At $3\times10^{-15}$~erg~s$^{-1}$~cm$^{-2}$ the simulated and the observed flux distribution are equal again, this flux value corresponds to a luminosity of $\log L_X = 29.2$~[erg~s$^{-1}$] at the distance of IRAS 20050+2720. 
\citet{2005ApJS..160..379F} derive a scaling law for the expected point source sensitivity of \emph{Chandra} based on the COUP observations. Scaled to the distance of IRAS 20050+2720 and assuming an average absorbing column density of $N_H=2\times 10^{22}$~cm$^{-2}$ (Sect.~\ref{sect:nh}) our detection limit in the center of the field of view, where the PSF is small, should be $\log L_X = 29$~[erg~s$^{-1}$]. We require at least 20 photons in a single observation to attempt spectral fitting, which corresponds to a luminosity $\log L_X = 30$~[erg~s$^{-1}$].

In the \emph{Spitzer} observations the mean 90\% differential completness limit over the observed field is 15.3, 15.2, 14.6, 13.6 and 8.8  mag for channel 3.6, 4.5, 5.8, 8.0 and 24\micron{} respectively. To derive the limits, artificial sources with different magnitudes are inserted into the observed mosaics. The limits give the faintest magnitude that allowed the pipeline to recover 90\% or more of the simulated sources \citep{2005ApJ...632..397G}. These limits vary over the field-of-view and they are on average less sensitive than in \citet{2009ApJS..184...18G} (who found 17.1, 16.8, 14.8, 13.8 and 7.0 mag). The new observations have similar exposure times, but the field of view now covers large regions outside the cloud, and thus without background extinction, so source confusion is a problem at  3.6 and 4.5\micron{}. However the density of 24~\micron{} sources outside the cloud is low, thus a lower fraction of simulated sources suffers from source confusion and the 90\% differential completness limit is at a fainter magnitude.
In the center of the cluster this work is more sensitive than \citet{2009ApJS..184...18G} because the new data reduction significantly enhances the sensitivity in the vicinity of bright sources (Sect.~\ref{sect:improvedspitzer}).

The sensitivity of the optical data is very non-uniform because of instrumental problems near chip edges and because of the wide and non-symmetric PSF, which prevents the detection of even moderately bright objects close to foreground stars. %The absence of a measurement in $UBVRI$ in the table~\ref{sourcelist} thus does not necessarily mean that the source is faint. 
The faint limit of the IPHAS catalog in $r'$ is 21~mag, its bright limit is 13~mag \citep{2005MNRAS.362..753D}. Thus at the distance of IRAS 20050+2720 the combination of FLWO photometry and IPHAS should be complete to about $r'=20$~mag, which corresponds to a mass of 0.1, 0.5 and 2.2~$M_{\sun}$ for $A_r$ 0, 5 and 8~mag, respectively, for a star of 1~Myr \citep{2000A&A...358..593S}.

The detection limits are very different from band to band. Also, reddening affects the optical observations much more than the IR data. Thus these samples have different detection biases and will identify a different fraction of the true number of YSOs in IRAS 20050+2720. 

\subsection{Sample summary}
After removal of foreground objects, the sample of YSOs in IRAS 20050+2720 now consists of 276 objects. Class I*, I, II and II* objects are classified purely by their IR properties. The main limitation for this dataset is source confusion in dense regions. All these sources may be detected in X-rays. We have identified the X-ray bright members of the class~III population, which we assume to be young, disk-less cluster members, and we will use the term ``class~III'' for this group in the remainder of the article for simplicity. This sample is complete to about $\log L_X = 29$~[erg~s$^{-1}$]. XYSOs are defined as an overdensity of X-ray sources without IR classification. Thus, by definition, isolated X-ray sources are not included. The XYSOs are associated with the densest part of the cluster and likely consist of embedded YSOs. %Strictly speaking, only Class I, II and II* are YSOs confirmed from the IR. Class I* and XYSOs are detected in too few bands to confirm their YSO nature. WTTS are YSO, but classified as from their X-ray emission.

In the following section we often split the sources into three groups called class I, II and III. In this case, class I includes class~I* and class~II includes class~II*. 

\section{Results}
\label{sect:results}
In this section we show our results. There is some overlap with \citet{2009ApJS..184...18G}, where some of the IR data are already presented. However, our work covers a larger region and wavelength range.

\subsection{Spatial distribution of sources}
\begin{figure}
\plotone{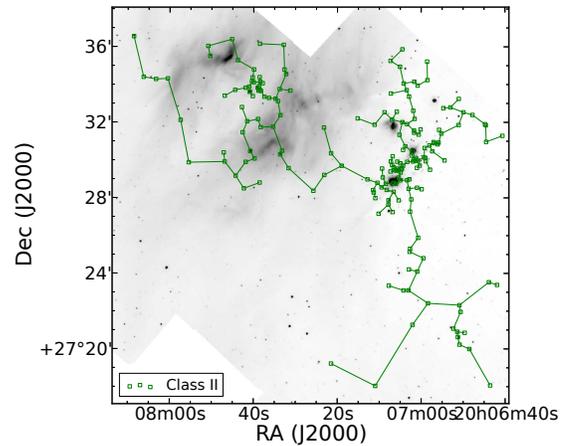}
\caption{MST for all class II sources, including transition disks. The sources (nodes of the graph) are marked with squares. The gray scale image shows IRAC 8~\micron{} emission.
See the electronic edition of the Journal for a color version 
of this figure.\label{fig:MST}}
\end{figure}

Here we characterize the spatial distribution of the sources using a minimum spanning tree analysis. A complete graph (i.e. each source is directly connected to every other source) is constructed, where the sources of interest are the ``nodes'' of the graph. For each connection between the sources (``edges of the graph'') we calculate the angular distance between the sources (``weight of an edge''). A subset of the connections is chosen in a way that minimizes the total length but keeps all sources connected.
This is called a minimum spanning tree (MST). As an example, Fig.~\ref{fig:MST} shows the MST for all class II sources.

\begin{figure}
\plotone{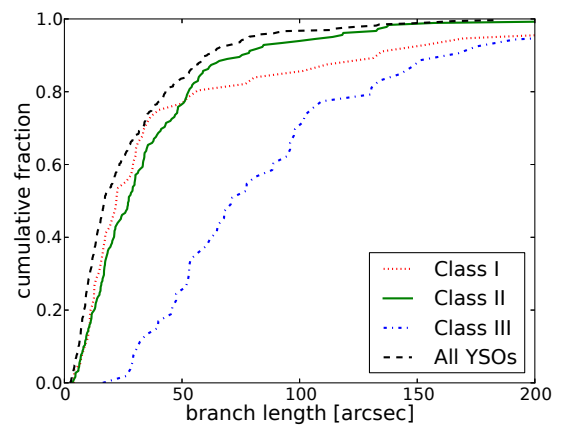}
\caption{Distribution of the branch length for different MSTs. 
See the electronic edition of the Journal for a color version 
of this figure.\label{fig:MSTbranchlength}}
\end{figure}

Fig.~\ref{fig:MSTbranchlength} shows the distribution of the branch length for MSTs of class I, II and III objects as well as the MST of all YSOs in the cluster. For class~I sources most branches are relatively short; the distribution for class~II sources is similar. In contrast, the mean branch length of class~III objects is much longer. Using an MST has some advantages to the more commonly employed nearest-neighbor (NN) distance and its variants (e.g. the distance to the fifth-nearest-neighbor - NN5) because the MST naturally results in connected clusters. Although both methods perform similarly on round shaped clusters, the MST is less susceptible to splitting elongated clusters. This is potentially an important drawback of NN methods for star formation in the elongated filaments of a molecular cloud.

One important restriction to keep in mind is that all these methods analyze the cluster as a two-dimensional structure on the plane of the sky. Thus we only see the projected separation of sources which is a lower limit to the real distance. Also, physically distinct subclusters may overlap when projected on the two-dimensional sky. In this case mathematical tools such as the MST or the NN method cannot provide a distinction between these two clusters.

The branch length distribution for class~III objects looks qualitatively different from the distribution of class~I and II sources. The break between two different linear regimes is less pronounced. Also, the average branch length is longer than for the other classes, which is partly due to the lower number of sources. There are 70, 188 and 54 class~I, II and III sources respectively. For a uniform distribution of sources over the same area we would expect that the branch length of the class~II sources is the smallest, and the other classes should have distances about a factor of two larger. Instead, class~I sources have the smallest branch lengths, thus they must be most clustered. Class~II sources are very similar and class~III objects have much larger branch lengths, indicating a more uniform distribution.

\begin{figure}
\plotone{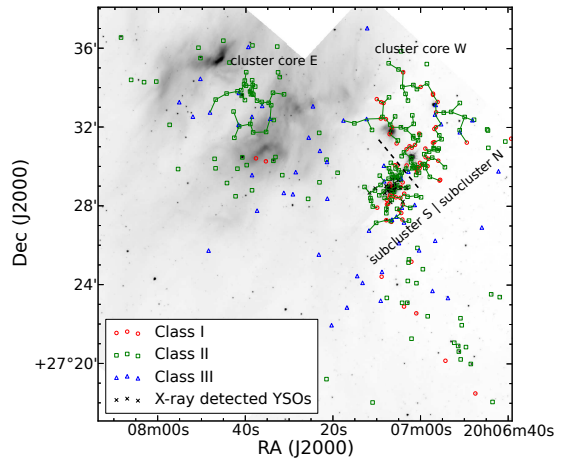}
\caption{Distribution of all YSOs in the sample. Members of the two cluster cores are connected by an MST. The gray scale image shows IRAC 8\micron{} emission.
See the electronic edition of the Journal for a color version 
of this figure. The black dashed line shows one (arbitrary) way to divide the western cluster core in two pieces, see text for discussion. \label{fig:MSTsubcores}}
\end{figure}

The upper and lower end of the branch length distribution for class~I and II sources can be fitted by two lines. The point of intersection for these lines can be used to algorithmically define a threshold branch length for all stars in a cluster. We cut all edges in the MST, which are longer than this threshold branch length of 45\arcsec{}, leaving several unconnected subclusters called ``trees''. \citet{2009ApJS..184...18G} introduced this technique in a comparison study of young clusters within 1~kpc of the sun and they show that it is well suited to a range of cluster morphologies.
Cutting all branches in the MST longer than the threshold branch lengths of 45\arcsec{} (0.15~pc), results in a minimum spanning ``forest''. The two biggest trees have 193 and 37 members, 6 trees have between 3 and 8 members. Fig.~\ref{fig:MSTsubcores} shows the distribution of all YSOs in the cluster. Overlaid are the MSTs for the two biggest cluster cores (``trees''), which are very different in their composition. The bigger cluster core on the western side of the cluster ($(\alpha, \delta)_{ {\rm J}2000}= 20^{\rm h}07^{\rm m}00^{\rm s},+27^\circ30'$, from here on cluster core W) has 175 members and a ratio of class I:II:III sources of 61:97:17. The smaller, eastern, one ($(\alpha, \delta)_{ {\rm J}2000}= 20^{\rm h}07^{\rm m}40^{\rm s},+27^\circ33'$, from here on cluster core E) has 37 members and a ratio of class I:II:III sources of 0:31:6, i.e. the ratio of class~II to III is very similar, but class~I sources are absent in the cluster core E. This is somewhat surprising given that cluster core E is surrounded by a reservoir of gas which can be seen in emission in 8\micron{} in Fig.~\ref{fig:MSTsubcores}. The class I:II:III ratios reported here are the detected ratios, which do not accurately represent the true ratios in the the cluster, because class I and II sources are detected from the IR, whereas class~III objects also require an X-ray detection and thus are subject to a different selection bias as explained in section~\ref{sect:completness}. The number of class~III objects in cluster core E is potentially underestimated, because it is further away from the \emph{Chandra} aimpoint than cluster core~W. Together with the lower number of class~0 sources this argues for a higher age.

Cluster core E is outside of the area surveyed by \citet{2009ApJS..184...18G}, but cluster core W is present in their analysis. Using the same technique they find a lower value for the threshold branch length (28\arcsec{}) and our cluster core W is divided into two parts, which they call Core-1 and Core-2. 
Even when we cut our MST at the same threshold branch length we still only recover a single core in the west; Core-1 and Core-2 appear united with the added YSOs reported here.

However, it is possible to separate cluster core W manually. This cut may or may not be physical. We present it here to discuss limitations of the MST technique. If we cut the cluster core W along the dashed black line (Fig.~\ref{fig:MSTsubcores}), we end up with two subcores. Their ratio of class~I:II:III is very similar with 21:41:12 and 29:54:7 for the northern and southern subcore of cluster core W, respectively. In the north there are more class~III objects, but, given the low numbers, this is only marginally significant. Both subcores differ significantly in the mean branch length. This is 18\arcsec{} in the northern subcore of cluster core W, but only 9.6\arcsec{} in the southern subcore of cluster core W. The MST analysis cannot distinguish if these are two separate subcores, which just happen to overlap in projection on the plane of the sky or if the northern subcore of cluster core W is just an extension of the southern subcore of cluster core W with lower stellar densities. If this is the same physical cluster core, than it is asymmetric extending only 2\arcmin{} towards the south of the densest core but covering a much larger area in the north.

In summary, we see that class~I sources are the most clustered and class~III objects are least clustered group in IRAS 20050+2720. The whole region consists of cluster core E with 37 members, that is surrounded by gas and dust, but has no class~I members and the much bigger cluster core W, which contains the densest and most extincted parts of IRAS 20050+2720 and has sources of all evolutionary stages.

Table~\ref{tab:clusterprops} describes the full MST and the MSTs for cluster cores W and E in order to facilitate a comparison with the sample of all young stellar clusters within 1~pc of the sun from \citet{2009ApJS..184...18G}. $R_{circ}$ is the radius of the smallest circle that contains all cluster members. Because most clusters are not circular we construct a convex hull for a cluster and define $R_{hull}$ as the radius of a circle, that has the same area as the convex hull, with an adjustment accounting for the number of stars that span up the hull \citep{2009ApJS..184...18G}. The aspect ratio is $R_{circ}/R_{hull}$. In addition, the average density of YSOs within the convex polygon $\sigma_{mean}$ is calculated and the peak density $\sigma_{peak}$ of the cluster after an adaptive smoothing \citep[for details see][]{2009ApJS..184...18G}. Last, the table shows the median nearest-neighboor distance.

\begin{table*}
\caption{\label{tab:clusterprops}Geometrical properties of IRAS 20050+2720 and its cluster cores.}
\begin{center}
\begin{tabular}{lcccccc}
\hline \hline
cluster & $R_{circ}$ & $R_{hull}$ & Aspect Ratio & $\sigma_{mean}$ & $\sigma_{peak}$ & Median NN2 length \\
&pc & pc & & pc$^{-2}$ &  pc$^{-2}$ & pc \\  \hline
full cluster & 1.8 &  1.37 &  1.71 &   60 & 5000 & 0.029 \\
cluster core W & 0.6 &  0.38 &  2.27 &  400 & 5000 & 0.020 \\
cluster core E&  0.3 &  0.22 &  1.36 &  200 & 1500 & 0.027 \\
\hline
\end{tabular}
\end{center}
\end{table*}

IRAS 20050+2720 was already one of the largest clusters in \citet{2009ApJS..184...18G}. Here, we have included the larger \emph{Spitzer} field of view and consequently, the size of the cluster has increased. Even cluster cores E and W separately have a large number of members compared to the clusters in the sample. Also, the mean and the peak densities are significantly higher than reported by \citet{2009ApJS..184...18G}. This is due to the added \emph{Chandra} data with allows us to identify cluster members by their X-ray emission. Since the distribution of class~III sources overlaps in space with class~I and II sources the average density is higher. The XYSOs found in the region with the highest extinction are closest together and thus define the peak density. This also leads to a very small median NN2 distance.

\subsection{Age of the cluster}
While \citet{1997ApJ...475..163C} estimate an average cluster age of 1~Myr, they interpret the molecular outflow and the stronger reddening in the central parts of subcluster W as signatures for a more recent star formation event in the last 0.1~Myr. Similarly, the ratio of sources in different stages suggests that cluster core W is younger than cluster core E. Further, cluster core W contains many class~I sources as well as a group of deeply embedded objects (XYSOs), both indicative of very young age.

\begin{figure}
\plotone{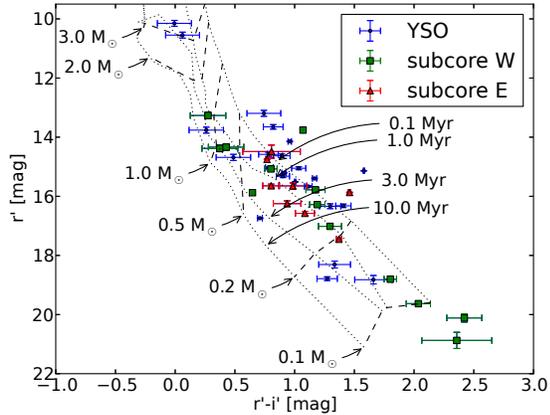}
\caption{CMD for all YSOs in our sample in SDSS bands. All stars are dereddened with the $A_K$ found during the source classification. The plotted errorbars represent only the photometric errors in $r'i'$ and do not include any statistic or systematic uncertainty in $A_K$. Overplotted are pre-main sequence isochrones from \citet{2000A&A...358..593S}.
See the electronic edition of the Journal for a color version 
of this figure.\label{fig:rriYSO}}
\end{figure}

Fig.~\ref{fig:rriYSO} shows a $r'$ vs. $(r'-i')$ CMD. It includes all YSOs with IPHAS detections and additional nine YSOs detected in $RI$ with FLWO. 
The IPHAS filters are similar, but not identical to, the SDSS filters. We use the following transformations which are appropriate for moderately red stars and have been tested on a G2V star (M. Guarcello, private communication): $r' = r'_{IPHAS} + 0.12$ and $i' = i'_{IPHAS} + 0.407 + 0.11 \times r'_{IPHAS}$.
Stars with $RI$ photometry from FWLO are converted to $ri$ based on the inverted relations from Lupton\footnote{http://www.sdss.org/dr7/algorithms/\\sdssUBVRITransform.html\#Lupton2005}.
% \begin{eqnarray}
% I & = & r - 0.2936 (r-i) - 0.1439\\
% R & = & r - 1.2444 (r-i) - 0.3820
% \end{eqnarray}
The transformation between $ri$ and $r'i'$ is done according to the formulas given on the SDSS website\footnote{http://www.sdss.org/dr7/algorithms/\\jeg\_photometric\_eq\_dr1.html\#usno2SDSS}. All stars are dereddened with the $A_K$ value, that was estimated during the source identification. The ratios of the reddending coefficients were taken from the compilation of \citet{1998ApJ...500..525S}.

The observed scatter in the CMD is, to some extent, due to instrumental uncertainties, but also to intrinsic scatter in the luminosity caused by binaries, variability and accretion. \citet{2005MNRAS.363.1389B} observed a time series of photometry for two OB associations and found that the temporal variability can account for up to half of the spread in their CMD. In the Upper Sco association \citet{2008ApJ...688..377S} show with Monte-Carlo simulations that all their observed scatter can be explained by variability, measurement uncertainty, and unresolved binaries despite the fact that spectroscopy is available to determine spectral types. Also, \citet{2008ASPC..384..200H} compared different sets of pre-main sequence evolutionary tracks and found significant offsets between the ages derived from them. Even worse, \citet{2010ApJ...722..971C} show an example where not even the relative age of different star forming regions is conserved when using different simulated tracks. They also highlight that most tracks do not include the additional luminosity from accretion spots, while this contributes to the observed photometry. This effect can shift ages by 50-100\% or introduce scatter of the same order, if different objects have varying accretion rates. All this will cause an apparent spread in luminosity. In addition, even a real spread in intrinsic luminosity does not require a spread in age. Instead, it can be caused by the individual history of strong accretion events for very young stars \citep{2011arXiv1102.4752J}.

Essentially all stars in Fig.~\ref{fig:rriYSO} are younger than 3 Myrs for cluster core W, possibly with an age spread. Cluster core E is compatible with an age below 1 Myr, although no class~I source is found there. If there is a real age spread in cluster core W, this can explain both the existance of very young and embedded objects like jet driving sources \citep{1995ApJ...445L..51B,1999A&A...343..585C,2008A&A...481...93B}, XYSOs, and class~I sources and also the older ages indicated in Fig.~\ref{fig:rriYSO}, since the younger members are likly too embedded to be observed in $r'i'$ and thus would not be shown in the figure.
An alternative explanation could be the presence of a mechanism that destroys disks, such as radiation and winds from a massive star.
However, no massive star is observed in this region; also, cluster core E is surrounded by much stronger gas emission than cluster core W, so any gas destruction would have to be confined to the disks, but not the surrounding material.

No systematic differences are seen between class~I and II objects, if they are plotted separately in the figure. 

We can confirm a cluster age of $<3$~Myrs for IRAS~20050+2720 and $<1$~Myr for cluster core E, although the errorbars in Fig.~\ref{fig:rriYSO} do not contain all sources of uncertainty. While we cannot distinguish between a past star formation event and ongoing star formation with certainty, the scatter in luminosity and thus apparent age is large and could indicate a real age spread in cluster core W.

\subsection{X-ray properties}

\subsubsection{Detection fraction}
\begin{figure}
\plotone{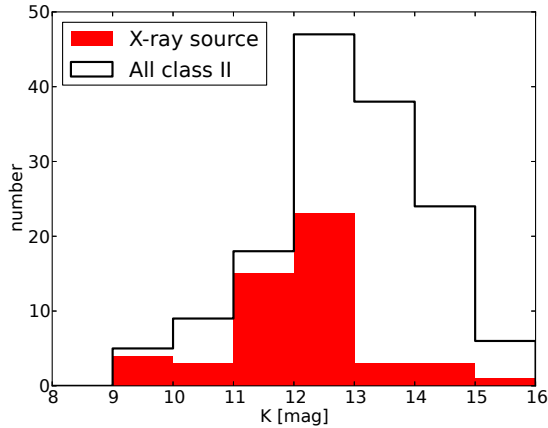}
\caption{Histogram of dereddened K band magnitude for class~II sources.
\label{fig:CIIwoXrays}}
\end{figure}

Table~\ref{tab:sources} shows the number of IR sources and X-ray sources in each class. Few class~I sources have X-ray detections, but for class~II sources we can compare the properties of those with and without an X-ray detection. 
Figure~\ref{fig:CIIwoXrays} shows their K band magnitude distribution. All values are dereddened with $A_K$ as found during the source classification. The mean value is $A_K=0.6$ over all class~II sources. X-ray bright sources are also bright in the 2MASS and \emph{Spitzer}/IRAC bands. Half of all sources with $m_K<12$~mag are X-ray sources but the detection fraction drops sharply for less luminous sources. In a magnitude limited sample with dereddened $m_K<13$~mag the $A_K$ values are fully compatible between X-ray detected and X-ray non-detected sources (KS-test probability that $A_K$ is drawn from the same sample $p=0.99$), but the magnitudes are not ($p=1^{-10}$). This indicates that the non-detection of X-rays is not caused by higher extinction but is intrinsic to the sources, i.e. objects fainter in the IR are weaker X-ray emitters and thus remain undetected.
According to pre-main sequence evolutionary tracks from \citet{2000A&A...358..593S} $m_K=12$~mag corresponds to stellar masses of $0.3$, $0.7$ and $0.9\;M_{\sun}$ at 1, 2 and 3~Myrs, respectively.

The regions of the highest absorbing column densities are all part of cluster core~W, thus this should show the lowest X-ray detection fraction, if the luminosities are uniform, contrary to the observed distribution. Also, stars in cluster core~W are brighter in K band than sources in the other regions. This might indicate an evolutionary trend: Cluster core~W still contains many class~I sources and also the class~II sources could on average be younger and less contracted. This would explain the higher K band luminosity and also the higher X-ray flux, since \citet{2010AJ....140..266W} show that the X-ray flux is linearly related to the surface area, such that larger YSOs have a higher luminosity.

\subsubsection{Absorbing column density}
\label{sect:nh}
\begin{figure}
\plotone{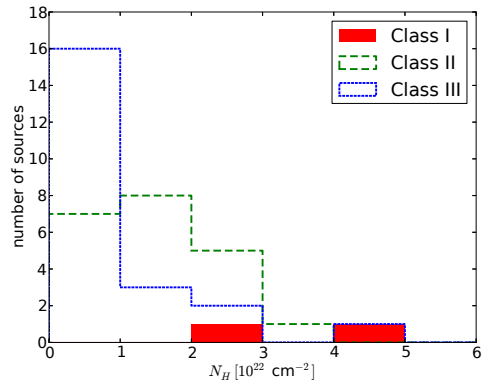}
\caption{Histogram of absorbing column densities $N_H$ for class~I, II and III sources.
See the electronic edition of the Journal for a color version 
of this figure.\label{fig:nh}}
\end{figure}

Some cluster members have been identified by their X-ray emission in Sect.~\ref{sect:xraysources}, but for an IR selected sample table~\ref{tab:sources} shows that the X-ray detection fraction increases with the evolutionary status of the sources from class~I* to class~II*. This is expected because the envelope of younger objects absorbs the soft X-rays and thus reduces the observable flux. 
In the following, we present results for fits to the X-ray spectra (see Sect.~\ref{sect:chandra}). Because of the low signal the fit does not converge for all sources and we ignore those fits, where the statistical error on the fit value includes the complete parameter space.
Fig.~\ref{fig:nh} confirms that class~I sources are indeed more embedded than class~II and III sources. According to a KS test, the probability that class~I and II sources are drawn from the same sample is 23\%. The probability that class~II and III sources are part of the same sample is only 3\%. This has also been observed in other star foming regions e.g. the ONC \citep{2008ApJ...677..401P} or NGC~1333 \citep{2010AJ....140..266W}.

While the deeply embedded sources may remain undetected, so that the figure does not represent the complete distribution of extinctions, the differences in the distribution of $N_H$ between different classes are real. Both class~I objects with fits have $N_H = 2-6 \times 10^{22}$~cm$^{-2}$, much greater than the average class~II or III source.
Less absorbed sources are easier to detect and if less absorbed class~I sources were present, they would have been detected. The lower detection fraction of younger sources in X-rays (table~\ref{tab:sources}) means, that class~I sources are fainter or they are so deeply embedded that the detections in Fig.~\ref{fig:nh} represent only the less absorbed end of the distribution. Thus, we conclude that sources with larger infrared excesses are indeed more absorbed in X-rays.

\subsubsection{Temperature}

\begin{figure}
\plotone{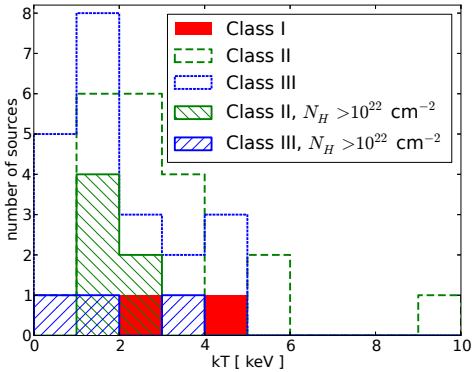}
\caption{Histogram of X-ray temperatures k$T$ for class~I, II and III sources. Sources with $N_H > 10^{22}$~cm$^{-2}$ are shown in hatched histograms. No separate histogram is plotted for class~I sources, because all of them show $N_H > 10^{22}$~cm$^{-2}$ anyway.
See the electronic edition of the Journal for a color version 
of this figure.\label{fig:kt}}
\end{figure}
Figure~\ref{fig:kt} shows the distribution of temperatures for the different classes (open histograms). Clearly, class~I sources are on average hotter than the other sources. However, this could be due to a detection bias: Class~I sources are generally more absorbed, thus soft X-ray emission remains undetected. To avoid this problem a subsample with similar $N_H$ can be used to compare the temperatures, because all sources suffer the same bias towards higher temperatures.
Unfortunately, the low number of sources requires a relatively broad range of $2 \times 10^{22}$~cm$^{-2} < N_H$ in order to retain a sufficient number of sources (hatched histograms). This interval contains both class~I sources. Although a KS test is not strictly valid for such small numbers, it indicates that the distribution of the plasma temperatures for class~II and III sources in this subsample differs from the distribution in the full sample, because lower temperature plasma remains undetected. The subsample contains 2 class~I sources, 6 class~II sources and 3 class~III sources.
Even in this small subsample a KS test shows that the three temperature distributions are incompatible. The class~I sources are hotter than the class~II sources and those in turn are hotter than the class~III soures. However, there may still be a bias, because class~II and III sources are mostly on the lower end of the $N_H$ range in the subsample. Due to the larger source numbers we can repeat this test for class~II and III sources in smaller $N_H$ bins. For $N_H = 0-1 \times 10^{22}$~cm$^{-2}$, $N_H = 1-2 \times 10^{22}$~cm$^{-2}$ and $N_H = 2-3 \times 10^{22}$~cm$^{-2}$ the mean temperature of the class~II sources is always higher than for class~III sources and the probabilities that the temperatures for the two classes are drawn for the same distribution are $<$1\%, 3\% and 15\%, respectively.

Thus, class~II sources in our sample seem to be hotter on average than class~III sources and it seems likely that class~I sources are even hotter than class~II sources. This has been seen before in other young clusters. While no significant trend is found in the ONC \citet{2008ApJ...677..401P} show in their Fig.~13 that class~II objects are marginally hotter than class~III sources. \citet{2010AJ....140..266W} studied Serpens and NGC~1333. In these two clusters, the class~III sources are found to have lower values of k$T$ than the class II.  Both samples are thought to have similar completeness for each class. The K-S probabilities that the class II and III temperature arise from the same distribution are 21\% (Serpens) and 3.3\% (NGC~1333). Our result agree with these findings, and the fact that we can show the effect in different $N_H$ ranges makes it very unlikely that this is caused by absorption only. However, the spectral models we use are based on a single temperature. This is an adequate description of the observed data with its low signal, but we know from observations of closer CTTS and WTTS that the real temperature structure in coronae is more complex. Also, the comparison in different $N_H$ bins could be flawed, if the $N_H$ is caused by a partial absorber in class~II or III sources, because only a part of the X-ray emission is covered by accretion funnels. Both considerations cannot be tested with the given data quality, thus we caution that the observed difference in temperature might not represent the real coronal temperature very well.

On first sight, the observed temperature trend seems to contradict the soft excess found in CTTS, mostly from \ion{O}{7} line emission \citep{RULup,manuelnh,2011AN....332..448G}, but \emph{Chandra}/ACIS is less sensitive at low energies than \emph{XMM-Newton} and only a very small absorbing column density ($N_H \approx 1 \times 10^{21}$~cm$^{-2}$) is needed to absorb the soft emission. Thus, the data in Fig.~\ref{fig:kt} show that the hot plasma component is hotter in class~I and II sources than in class~III sources, but the data do not constrain the presence of a soft emission component from accretion shocks. 

\subsubsection{Luminosity}
In clusters of stars there is a wide distribution of X-ray luminosities. Luminosities for stars with spectral fits are given in table~\ref{tab:xray}, but this applies only to a fraction of the YSOs with X-ray detections. To estimate a rough luminosity for the remaining sources, we fit a relation between count rate $c$ and computed unabsorbed source flux $f$ to the YSOs with spectral fits. We find the following relation: $f = c * (4\pm2)\times 10^{-9}$~erg~s$^{-1}$~cm$^{-2}$~count$^{-1}$. While this is very uncertain for a single star because the proportionality constant really depends on the temperature and absorption, this fit is a reasonable approximation for a sample of stars, because the relation represents the average values of $N_H$ and k$T$ for those YSOs with spectral fits. Figure~\ref{fig:CIIwoXrays} indicates that the X-ray detected sources, at least for class~II objects, are the brighter in K-band and thus are likely more massive stars. In this case, the sources with weak X-ray emission might just be the less massive and thus intrinsically fainter than their counterparts with X-ray spectral models, but could still have similar $N_H$ and k$T$ values.

\begin{figure}
\plotone{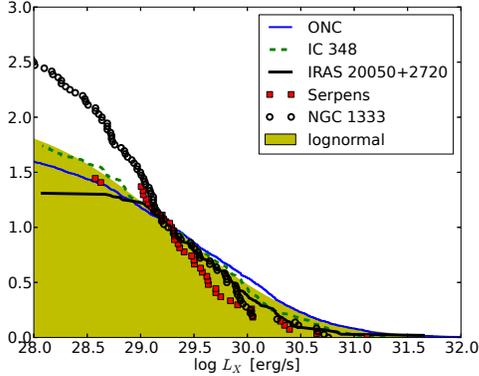}
\caption{The plot shows cumulative X-ray luminosity functions (XLF) for different star forming regions normalized to 1.0 at $\log L_X = 29.3$. The data for Serpens and NGC~1333, which have less than 100 measured values, are shown as individual points. The shaded area shows an analytical lognormal distribution with $\mu = 29.3$ and $\sigma = 1.0$ as suggested by \citet{2005ApJS..160..379F}. See text for data sources.
See the electronic edition of the Journal for a color version 
of this figure.\label{fig:XLF}}
\end{figure}

Above $\log L_X = 29.8$ the X-ray luminosity function (XLF) is dominated by sources with spectral fits, thus the sample is fairly complete and we are confident in the $L_X$ values. However, soft sources could be hidden in the dense parts of the cloud, even if they have a high $L_X$. Most sources below this value have luminosities which are only estimated from the count rate.
Figure~\ref{fig:XLF} shows the cumulative X-ray luminosity function for IRAS 20050+2720 and several other star forming regions for comparison: The Orion nebula cluster \citep[ONC,][]{2005ApJS..160..319G}, IC~348 \citep{2002AJ....123.1613P}, and Serpens and NGC~1333 \citep{2010AJ....140..266W}. 

\citet{2005ApJS..160..379F} suggested that the XLF of young star forming regions could be universally described by a lognormal distribution with $\mu = 29.3$ and $\sigma = 1.0$. In Fig.~\ref{fig:XLF} The remaining clusters in the figure are complete only for higher luminosities. The high luminosity end of the distribution is defined by very few massive O and B stars, which are not present in the lower mass clusters. The ONC follows a lognormal curve very well here. 
Differences on the high-mass end of the XLF are seen between the ONC and NGC~2244, a similarly massive cluster \citep{2008ApJ...675..464W}, and there seems to be an excess of sources around $\log L_X = 29.3$ in Cep~B \citep{ 2006ApJS..163..306G}. Here we compare it to several clusters of lower mass.
The ONC sample should be nearly complete down to $M_*<0.2M_{\sun}$. Serpens and NGC~1333 have the steepest XLF and the lowest cluster mass ($<200\;M_{\sun}$), IC~348 and IRAS 20050+2720 \citep[300 and 430~$M_{\sun}$,][]{2009ApJS..184...18G} have a shallower XLF and intermediate masses and the ONC, the most massive cluster in the figure, has the shallowest slope and follows a lognormal distribution with $\sigma = 1.0$ most closely. We compare the XLF above $\log L_X = 29.8$ with 2-sided KS tests. In Serpens there are only 10 sources in this range, thus it is compatible with any other distribution. Apart from Serpens, the chance that the ONC sample and any other XLF are drawn from the same parent distribution is $<2$\%. IRAS~20050+2720 agrees with IC~348 and NGC~1333 on the 30\% level, while the chance that IC~348 and NGC~1333 are drawn from the same parent sample is 19\%. Thus, we see a devitation of the observed XLFs for low-mass clusters from the COUP XLF, a further indication that the XLF is not universal.

The more massive clusters contain a larger number of massive stars of the spectral types B to F, which are by themselves relatively faint in X-rays. However, a high fraction of them are binaries with an X-ray bright companion, i.e. a G or K type star. These binaries could provide additional X-ray luminosity in the range $\log L_X = 29-30$ and thus tilt the observed distribution to explain why more massive clusters have a shallower XLF.

\subsection{Gas absorption and dust extinction}

The absorption of X-rays is mainly due to metal atoms and ions in the line of sight. It is generally expressed as the equivalent hydrogen column density $N_H$. In contrast, optical and infrared extinction is mainly caused by dust. Simplistically, it should be possible to calculate the gas-to-dust ratio along the line of sight for those sources with an X-ray fit and sufficient IR information to estimate $A_K$. However, the situation is complicated by grain coagulation and ice growth. 

\begin{figure}
\plotone{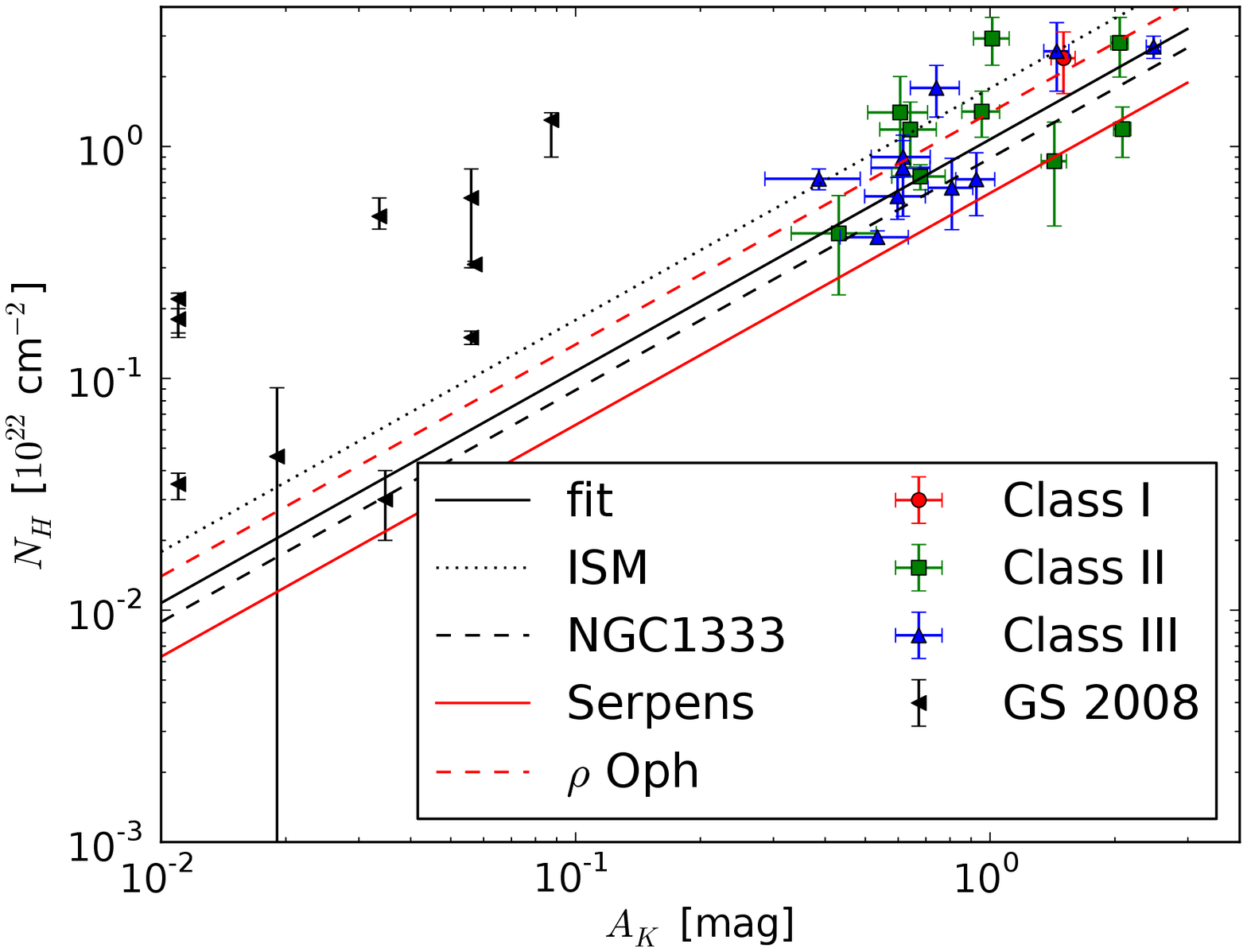}
\caption{Relation between gas column density $N_H$ and extinction $A_K$. Different symbols show different source types for our sample. For comparison, all CTTS from \citet{FUSElineforms} are also shown. Lines indicate a fit to the data for IRAS 20050+2720, the interstellar gas-to-dust ratio from \citet{2003A&A...408..581V} and the ratio found in NGC~1333 by \citet{2010AJ....140..266W}. All lines have the same slope in this log-log plot, because they are all derived assuming $N_H \propto A_K$ and only differ in the proportionality constant. 
See the electronic edition of the Journal for a color version 
of this figure.\label{fig:aknh}}
\end{figure}

Figure~\ref{fig:aknh} shows the $N_H$ vs. $A_K$ for IRAS 20050+2720 with different symbols for different evolutionary classes. Due to the uncertainty in the intrinsic color of the stars, there is an uncertainty of 0.1 in $A_K$ (R. Gutermuth, personal communication).
A linear fit gives the relation $N_H = (1.1 \pm 0.1) \times 10^{22}A_K$~cm$^{-2}$~mag$^{-1}$. There is no significant difference between class~II and III sources and separate fits for cluster core E and W lead to compatible values. The $N_H/A_K$ ratio is lower than the interstellar ratio of $N_H = 1.8 \times 10^{22}A_K$~cm$^{-2}$~mag$^{-1}$ \citep{1995A&A...293..889P}, where $A_V$ was transformed to $A_K$ with the tabulated $A_K/A_V$ of \citet{1998ApJ...500..525S}. However, values lower than in IRAS 20050+2720 are sometimes seen in young star forming regions, e.g. $N_H = (0.89 \pm 0.13) \times 10^{22}A_K$~cm$^{-2}$~mag$^{-1}$ for class~III sources in \object{NGC 1333} \citep{2010AJ....140..266W} and $N_H = (0.63 \pm 0.23) \times 10^{22}A_K$~cm$^{-2}$~mag$^{-1}$ for class~III sources in Serpens. In the $\rho$~Oph cloud \citet{2003A&A...408..581V} find $1.4\pm0.1\times 10^{22}A_K$~cm$^{-2}$~mag$^{-1}$.

The situation appears more consistent with the ISM in higher mass star forming regions.
In \object{RCW 108} (with at least one O8 star) \citet{2008AJ....135..693W} obtained
$N_H = (1.6-2.60) \times 10^{22}A_K$~cm$^{-2}$~mag$^{-1}$ for different subregions of the cluster. For the O5 bearing star forming region RCW~38 \citet{2006AJ....132.1100W} find $N_H = 1.8 \times 10^{22}A_K$~cm$^{-2}$~mag$^{-1}$, again in agreement with the interstellar ratio.

In an effort to understand this bifurcation, figure~\ref{fig:aknh} also shows $N_H$ vs. $A_K$ for the very different regime of close-by and little absorbed class~II sources from the sample of \citet[][see references therein for data sources]{FUSElineforms}. Most of those sources have been observed with X-ray grating spectroscopy, thus the $N_H$ values are more reliable even for small column densities. On the other hand, the $A_V$ values for this sample have been collected from the literature and are derived with different methods. In the plot sources with a literature value of $A_V = 0$ are shown at $A_K=0.01$~mag, because it is difficult to measure a small reddening. Most of these sources are found significantly above the $N_H/A_K$ of the diffuse ISM. 

\citet{2007ApJ...669..493W} proposed  three possible reasons for the low $N_H/A_K$ values.  First, the X-ray temperatures could be systematically underestimated resulting in an over estimate of the $N_H$ values. Only X-ray grating spectroscopy can break the $N_H$ to k$T$ degeneracy in the fits. However, this is not feasible for class II or III sources beyond about 150~pc. Second, that the value of $A_K$ has been overestimated due to changes in the form of the reddening law; however, the required change would be much greater than has so far been observed. The third reason was that the growth and coagulation of grains would increase the extinction per unit grain mass in the cloud. They find the first two reasons unlikely, and argue that grain growth is the most likely explanation. 

While both $N_H$ and $A_K$ are measures of the material along the line of sight, $N_H$ is essentially a measure of number of atoms while extinction is affected by particle shape and the grain size distribution. In the case of a star, which is surrounded by a disk and has a gas accretion column, there are various
layers which can have different particle shapes and grain size distributions. Nonetheless, in IRAS 20050+2720 class~II and III sources show the same $N_H/A_K$ ratio, although class~II sources have more circumstellar material. Thus, either the extinction laws of the circumstellar material and the cloud are similar or the circumstellar extinction is not a
significant contributor to the total extinction. 

If we view $N_H/A_K$ as a gas-to-dust ratio, IRAS 20050+2720, Serpens and NGC 1333 are dust rich (or gas poor), compared with the interstellar extinction. In contrast, the sample from \citet{FUSElineforms} shows most sources to be dust depleted (or gas enriched) with much lower absolute values for the extinction. These sources are mostly from the Taurus-Auriga molecular cloud complex, which does not have the high density cluster cores seen in IRAS 20050+2720, Serpens and NGC~1333, and contributes little to the line-of-sight extinction. Thus, their extinction should be dominated by circumstellar material, which could be dust-depleted due to the energy input from the star, gravitational settling of grains in the disk-midplane and planet formation. %In contrast, in the younger star forming regions, we mostly see the cloud material, which could be dust-rich for those clouds without high-mass stars or metal poor in the gas phase due to grain formation. 
% 
% 
% 
% Alternatively, the apparent change in $N_H/A_K$ could be caused by a change in the dust grain size distribution. In this scenario the dust grows in the moderate density environments of low-mass star forming regions. 
% The $A_K$ is derived from the color excess in two bands, e.g. $E(J-K)$ using the ratio of the total to selective extinction $R_K = \frac{A_K}{E(J-K)}$. Larger grains scatter more light, thus extinction per unit mass could be increased in star forming regions, so that the same color excess would be interpreted as a different $A_K$ and thus we would derive a different $N_H/A_K$ ratio. 
%The interpretation of the larger $N_H/A_K$  ratio in older stars in Taurus-Auriga \citet{FUSElineforms} is some of the grains have grown to sizes in the mm range, too large to cause substantial reddening.  On the other hand, RCW~108 and RCW~38 are massive and dense star forming regions, their members could supply sufficient UV flux to destroy dust or inhibit grain growth in large parts of the cloud \citep{1999MNRAS.303..367S}, leading to an extinction law closer to the interstellar one.

In contrast, in the younger star forming regions, we mostly see the cloud material. Although we cannot rule out that these clouds could be dust-rich (perhaps from a high abundance of carbon) or poor
in X-ray absorbing metals, there is no independent evidence for these explanations. Alternatively, grain growth and coagulation can lead to an increase in the extinction per Hydrogen atom in the near IR-bands  \citep{1980ApJ...235...63J,2001ApJ...548..296W}. \citet{2007ApJ...666L..73C}  report a decrease in the ratio of the depth of the 9.7~$\mu$m silicate feature to the color excess $E(J-K)$ in molecular clouds relative to that in the diffuse ISM. They note that models predict grain growth can increase the amount of extinction per Hydrogen atom in the near-IR without
strongly affecting the shape of the reddening curve \citep{1994A&A...291..943O}; note that the value of $A_K$ is calculated primarily with the $E(J-H)$ color excess and $E(H-K)$ color excesses; these should follow a similar trend as the $E(J-K)$ color excess. Models of the grain coagulation in molecular clouds by \citet{2011A&A...532A..43O} show the $E(J-K)$ per unit column density can increase relative to the diffuse ISM due to coagulation until the grains start exceeding 1~$\mu$m in size. 

In this scenario, the low $N_H/A_K$ ratio is due to grain growth in the cold, dense molecular cloud material surrounding low-mass star forming regions; although there is some dispersion in the amount of grain growth since IRAS~20050+2720 shows a higher $N_H/A_K$ than NGC~1333 or Serpens. In contrast, the extinctions and gas column densities for class~II objects in the Taurus-Auriga region may have a significant component from the circumstellar disk; grains in these disks may have grown to sizes in the mm range, too large to cause substantial reddening, thus the \citet{FUSElineforms} measurements show a larger $N_H/A_K$ ratio. Alternatively, scattered light from the grains may also contribute to the colors of the objects causing an underestimate of the extinction. The RCW~108 and RCW~38 are massive star forming regions, whose members could supply sufficient radiant flux and/or mechanical power to evaporate grain mantles and break up coagulated dust in large parts of the cloud \citep{1999MNRAS.303..367S}, leading to an extinction law closer to the interstellar one.

Alternatively, \citet{2003A&A...408..581V} show that the low $N_H/A_K$ ratio in $\rho$~Oph, which is compatible with the value found in IRAS20050+2720, can be described by a change in the local abundance, that is consistent with recent revisions of the local (solar) abundance (see their paper for a detailed comparison of different sets of abundances). However, this would require a different set of abundances for each of $\rho$~Oph and IRAS~20050+2720, Serpens, and NGC~1333, while RCW~108 and RCW~38 both should have the same abundances as the galactic ISM. The abundances in the stellar photosphere have not been measured to the precision required to compare the different clusters, thus this idea cannot be confirmed with the available data.

In summary, the cold molecular clouds of low-mass star forming regions have $N_H/A_K$ ratios that are systematically lower than that of the diffuse ISM due to their cold, dense ($>10^4$~cm$^{-3}$) and relatively unperturbed environments which facilitates the growth of grains.

% \subsection{Radio properties}
% Only three IR sources are detected in the radio observation. The first one is an XYSO (ID 4), the second one a transition disk object (class II*, ID 274) and the third one is detected in three IRAS bands only (ID 20516), thus it cannot be classified, but its colors are compatible with a YSO. The peak fluxes are 160, 720 and 100~$\mu$Jy with an uncertainty of 50~$\mu$Jy. The region covered by the radio observations is much smaller than in the other wavelength bands, thus the numbers are too small to derive meaningful detection ratios. The detected objects show similar fluxes as the radio sources in the Coronet cluster, a star forming region in Corona Australis at 150~pc \citep{2006A&A...446..155F,2007A&A...464.1003F}. However, the distance to IRAS~20050+2720 is about 5 times larger and only the brightest source in the Coronet would have been found in the observation of IRAS~20050+2720, thus, the low number of detections is not surprising. This also highlights, that the transition disk object ID~274 is exceptionally radio bright, at least during the VLA observations.

\section{Summary}
\label{sect:summary}
We present multi-wavelength data for the young star forming region IRAS~20050+2720 from the \emph{Spitzer} and \emph{Chandra} satellites and the ground-based 2MASS and IPHAS surveys as well as targeted FLWO photometry. YSOs of class~I and II are identified based on their IR properties. X-ray sources with optical or IR counterparts that are compatible with the cluster distance are added to this sample as well as a group of X-ray sources that are identified as cluster members based on their close association with the densest cluster core.

We use a minimum spanning tree (MST) analysis to characterize the spatial distribution of sources in IRAS 20050+2720. Overall, class~I sources are more clustered than class~II sources, and class~II sources are more clustered than class~III sources. %Within the class~II sources those with an X-ray detection are more clustered (with $1\sigma$ confidence).

Two main cluster cores can be identified in IRAS 20050+2720. Cluster core~E consists only of class~II and III sources, but seems younger in the $r'$ vs. $r'-i'$ diagram. It is surrounded by an IR nebulosity.  Cluster core~W has a dense and highly extincted core in the South and a much less dense region in the north. It is not clear if those two regions are distinct groupings along overlapping lines-of-sight, or if they are part of the same physical cluster core. The age of IRAS 20050+2720 is $<3$~Myrs and it might contain populations of different ages.

In X-rays we detect preferentially the more massive YSOs; the detection fraction is higher for more evolved evolutionary stages, where the absorbing column densities are lower. Still, the detected class~I objects are both hotter and more luminous than class~II sources with similar absorption. We compare the XLF with other clusters and find an apparent correlation with cluster mass: The XLF can always be approximated by a lognormal distribution, but it seems to be steeper for clusters of lower mass.

Using the combined X-ray and infrared data, we derive the $N_H/A_K$ values for the YSOs in IRAS 20050+2720.  We find a value which is 60\% of the value for the diffuse ISM with no significant difference in this ratio between the cluster cores of IRAS 20050+2720 or between sources at different evolutionary stages. Compared to values derived with other YSOs
the $N_H/A_K$ ratio is higher than that for the low-mass regions Serpens or NGC~1333, but much lower than that observed for class~II objects in the Taurus molecular cloud where the extinction and X-ray absorption may be dominated by circumstellar material. This could be due to modest grain growth in the molecular cloud environment that dominates the extinction
toward IRAS 20050+2720; in contrast, the circumstellar environments may have elevated values of $N_H/A_K$ due to grain growth, gravitational settling, planet formation or dust evaporation.

\acknowledgements 
This work is based in part on observations made with the Spitzer Space Telescope, which is operated by the Jet Propulsion Laboratory, California Institute of Technology under a contract with NASA.
This publication makes use of data products from IPHAS (carried out at the Isaac Newton Telescope, INT, operated on the island of La Palma by the Isaac Newton Group) and from the Two Micron All Sky Survey, which is a joint project of the University of Massachusetts and the Infrared Processing and Analysis Center/California Institute of Technology, funded by the National Aeronautics and Space Administration and the National Science Foundation. PyRAF is a product of the Space Telescope Science Institute, which is operated by AURA for NASA. This work was funded by Chandra award GO6-7017X.

{\it Facilities:} \facility{CXO (ACIS)}, \facility{Spitzer (IRAC, MIPS)}, \facility{FLWO:1.2m (Keplercam)}


%\appendix
% 
% \bibliographystyle{../AAStex/astronat/apj/apj}
% \bibliography{../articles}

\end{document}